\documentclass{article}
\newcommand{\be}{\begin{equation}}
\newcommand{\ee}{\end{equation}}
\newcommand{\ba}{\begin{eqnarray}}
\newcommand{\ea}{\end{eqnarray}}

\def\vf{\varphi}
\def\g{\gamma}

\def\j{\psi}

\def\l{\lambda}
\def\m{\mu}

\def\p{\pi}
\def\q{\theta}

\def\x{\xi}
\def\z{\zeta}
\def\D{\Delta}

\def\cv{{\cal V}}

\newcommand{\ov}{\overline}

\newcommand{\aand}{\;\;\;\mbox{and}\;\;\;}
\newcommand{\pa}{\partial}
\newcommand{\sx}{\sigma_x}
\newcommand{\sy}{\sigma_y}
\newcommand{\sz}{\sigma_z}
\def\interior{\mathaccent "7017 }
\def\sl#1{\rlap{\hbox{$\mskip 1 mu /$}}#1}
\def\Sl#1{\rlap{\hbox{$\mskip 3 mu /$}}#1}
\def\SL#1{\rlap{\hbox{$\mskip 4.5 mu /$}}#1}
\def\I{\leavevmode\hbox{\small1\kern-3.8pt\normalsize1}}

\newcommand{\pari}{\stackrel{{P}}\longrightarrow}

\date{}

\title{{\bf Low-energy electron-electron bound states in planar QED}}

\author{Hugo R. Christiansen$^{1,2}$, Oswaldo M. Del Cima$^1$\footnote{Talk 
given at the \it{XXI Encontro Nacional de F\'\i sica de Part\'\i culas e Campos, 
October 2000, S\~ao Louren\c co - MG - Brazil.}} \footnote{The author dedicates this work 
to his daughter Vittoria who attended the talk.}, \\
Manoel M. Ferreira Jr$^{2,3}$ and Jos\'e A. Helay\"el-Neto$^{1,2}$ \\
\small\it $^1$Universidade Cat\'olica de Petr\'opolis (UCP), \\ 
\small\it Grupo de F\'\i sica Te\'orica (GFT), \\ 
\small\it Rua Bar\~ao do Amazonas 124 - 25685-070 - Petr\'opolis - RJ - Brazil. \\ 
\small\it $^2$Centro Brasileiro de Pesquisas F\'\i sicas (CBPF),\\
\small\it Coordena\c c\~ao de Teoria de Campos e Part\'\i culas (CCP), \\
\small\it Rua Dr. Xavier Sigaud 150 - 22290-180 - Rio de Janeiro - RJ - Brazil. \\  
\small\it $^3$Departamento de F\'\i sica,\\
\small\it Universidade Federal do Maranh\~ao (UFMA), \\
\small\it Campus Universit\'ario do Bacanga - 65085-580 - S\~ao Luiz - MA - Brazil.}

\begin{document}

\maketitle

\begin{abstract}
In this talk, we present a parity-preserving QED$_{3}$ model with spontaneous
breaking of a local $U(1)$-symmetry. The breaking is accomplished by a
potential of the $\vf^6$-type. It is shown that a net attractive 
interaction appears in the M{\o}ller scattering ($s$- and $p$-wave scatterings between two 
electrons) as mediated by the gauge field and a Higgs scalar. 
We show, by solving numerically the Schr\"odinger equation for both the scattering 
potentials ($s$- and $p$-wave), that in the weak-coupling regime 
only $s$-wave bound states appear, whereas in the strong-coupling regime 
$s$- and $p$-wave bound states show up. 
Also, we discuss possible applications of the model to the phenomenology of 
high-$T_c$ superconductors \cite{burns} and to the re-entrant 
superconductivity effect \cite{boebinger}.

\end{abstract}

\section{The planar QED and the Higgs mechanism}
The action for the parity-preserving QED$_{3}$\footnote{The metric is $\eta_{mn}$$=$$(+,-,-)$; %%@
$m$,
$n$$=$$(0,1,2)$, and the $\g$-matrices are taken as $\g^m$$=$$(\sx,i\sy,-i\sz)$.} 
with spontaneous symmetry breaking of a local $U(1)$-symmetry is given by \cite{delcima1}:
\ba
S^{\rm inv}_{{\rm QED}}\!\!\!\!&=&\!\!\!\!\int{d^3 x}     
\biggl\{ -{1\over4}
F^{mn}F_{mn}
+ i {\ov\j _+} {\SL{D}} {\j}_+ + i
{\ov\j _-} {\SL{D}} {\j}_- - m_e(\ov\j_+\j_+ - 
\ov\j_-\j_-) +\nonumber\\
\!\!\!\!&-&\!\!\!\! \l_{ep} (\ov\j_+\j_+ - 
\ov\j_-\j_-)\vf^*\vf + D^m\vf^* D_m\vf - V(\vf^*\vf)\biggr\}~,\label{inv}
\ea 
with the potential $V(\vf^*\vf)$ taken as
\be
V(\vf^*\vf)=\m^2\vf^*\vf + {\z\over2}(\vf^*\vf)^2 + {\l\over3}(\vf^*\vf)^3~, 
\ee
where the mass dimensions of the parameters, 
$\m$, $\z$, $\l$ and $\l_{ep}$ are
respectively ${1}$, ${1}$, ${0}$ and ${0}$.

The covariant derivatives are defined as follows:
\be
{\SL{D}}\j_{\pm}\equiv(\sl{\pa} + i\frac{e}{\sqrt\l_c} \Sl{A})\j_{\pm} 
\aand
D_{m}\vf\equiv(\pa_{m} + i\frac{e}{\sqrt\l_c} A_{m})\vf~,\nonumber 
\ee 
where $\frac{e}{\sqrt\l_c}$ is a coupling constant with dimension of 
(mass)$^{1\over2}$ - the electron charge $e$ is dimensionless. In the
action (\ref{inv}), $F_{mn}$ is the usual field
strength for $A_m$, $\j_+$ and $\j_-$ are two kinds 
of fermions (the $\pm$ subscripts refer to their spin sign 
\cite{binegar}) and $\vf$ is a complex scalar. 

The QED$_{3}$-action is invariant under the parity 
symmetry, $P$, whose action is fixed below:
\ba
x_m &\pari& x_m^P=(x_0,-x_1,x_2)~,\nonumber\\
\j_{\pm} &\pari& \j_{\pm}^P=-i\g^1\j_{\mp}~,~
\ov\j_{\pm} ~\pari ~\ov\j_{\pm}^P=i\ov\j_{\mp}\g^1~,\nonumber\\
A_m &\pari& A_m^P=(A_0,-A_1,A_2)~,\nonumber\\
\vf &\pari& \vf^P=\vf~.
\ea 

The sixth-power potential, $V(\vf^*\vf)$, is the responsible 
for breaking the electromagnetic $U(1)$-symmetry. It is the 
most general renormalizable potential in three dimensions.

Analyzing the potential $V(\vf^*\vf)$, and 
imposing that it is bounded from
below and yields only stable vacua (metastability is 
ruled out), the following conditions on the parameters $\m$, $\z$, $\l$ must be set :
\be
\l>0 ~,~ \z<0 \!\!\aand\!\! \m^2 \leq {3\over 16} 
{\z^2\over \l}~.
\ee 
We denote ${\langle}\vf{\rangle}$$=$$v$ and the vacuum 
expectation value for
the $\vf^*\vf$-product, $v^2$, is chosen as
\be
{\langle}\vf^*\vf{\rangle}=v^2=-{\z \over 2\l}+ 
\left[ \biggl({\z \over
2\l}\biggr)^2 - {\m^2\over \l} \right]^{1\over 2}~, 
\ee 
the condition for the minimum leading as $\m^2$$+$${\z}v^2$$+$${\l}v^4$$=$$0$.
%as below:
%\[
%\m^2+{\z}v^2+{\l}v^4=0~. 
%\] 

The complex scalar, $\vf$, is parametrized by $\vf$$=$$v$$+$$H$$+$$i\q$, 
%\[
%\vf=v+H+i\q~, \label{para}
%\] 
where $\q$ is the would-be Goldstone boson and $H$ 
is the Higgs scalar, both with vanishing vacuum expectation values.  

In order to preserve the manifest renormalizability 
of the model, the 't Hooft gauge is adopted:
\ba
{\hat S}_{R_{\x}}^{\rm gf}\!\!\!\!&=&\!\!\!\!
\int{d^3 x}\left\{
-{1\over 2\x} \biggl( \pa^m A_m -{\sqrt{2}}\x 
M_A \q \biggl)^2 \right\}~,
\ea 
where $\x$ is a dimensionless gauge parameter.

Replacing the field parametrization for $\vf$ 
into the action (\ref{inv}), and adding up the 't Hooft gauge, yields the following complete %%@
parity-preserving action:
\ba
S_{\rm QED}\!\!\!\!&=&\!\!\!\!\int{d^3 x}      
\biggl\{ -{1\over4}F^{mn}F_{mn} + {1\over2} M^2_A A^m A_m 
-{1\over 2\x}(\pa^m A_m)^2 + \nonumber \\
\!\!\!\!&+&\!\!\!\! i {\ov\j _+} {\SL{D}} {\j}_+ + i
{\ov\j _-} {\SL{D}} {\j}_- - m_e(\ov\j_+\j_+ - \ov\j_-\j_-) + 
\nonumber\\
\!\!\!\!&+&\!\!\!\!\pa^m H \pa_m H +
{\pa^m}\q {\pa_m}\q - \x M^2_A\q^2 + \nonumber\\
\!\!\!\!&-&\!\!\!\!
\l_{ep} (\ov\j_+\j_+ - \ov\j_-\j_-)((v+H)^2+\q^2) + 2\frac{e}{\sqrt\l_c}A^m(H{\pa_m}\q - %%@
\q{\pa_m}H) + 
\nonumber\\
\!\!\!\!&+&\!\!\!\! \frac{e^2}{\l_c} A^m 
A_m(2vH+H^2+\q^2) - \m^2((v+H)^2+\q^2) + \nonumber\\   
\!\!\!\!&-&\!\!\!\! {\z\over2}((v+H)^2+\q^2)^2 - {\l\over3}((v+H)^2+\q^2)^3 \biggr\}~.
\ea 
where the physical mass parameters $M^2_A$, $m$ and $M^2_H$ are 
given by
\be
M^2_A=2v^2\frac{e^2}{\l_c}~,~m=m_e+\l_{ep}v^2 \!\!\aand\!\!
M^2_H=2v^2(\z+2 \l v^2)~.
\ee

The M\o ller scattering to be contemplated will 
include the scatterings
mediated by the gauge field and the Higgs ($A_m$ 
and $H$). The scattered
electrons may exhibit either opposite spin polarizations 
($e_{(\pm)}^-$ and $e_{(\mp)}^-$) or the same spin 
polarizations ($e_{(\pm)}^-$ and $e_{(\pm)}^-$).
To compute the scattering potentials for the interaction 
between electrons with opposite spin polarizations ($s$-wave) 
and with the same spin polarizations 
($p$-wave), we refer to the works of \cite{sucher}, where the concept of potential 
in quantum field theory and in scattering processes is discussed in detail.

The calculation of scattering potentials will be performed in 
the center-of-mass frame, for in this frame the scattered electrons 
are correlated in momentum space.

The $s$- and $p$-wave scattering potentials turn out to be:
\ba
{\cv}^{s}(r)\!\!\!\!&=&\!\!\!\!-\frac{1}{2\p}\left[2\l_{ep}^2v^2 K_0(M_Hr)+
\frac{e^2}{\l_c} K_0(M_Ar)\right]~,\nonumber\\
{\cv}^{p}(r)\!\!\!\!&=&\!\!\!\!-\frac{1}{2\p}\left[2\l_{ep}^2v^2 K_0(M_Hr)- 
\frac{e^2}{\l_c} K_0(M_Ar)\right]~.
\ea 

Now, a particular condition on the parameters is set:
\be
\frac{e^2}{\l_c} = \z + 2\l v^2~~(M_H=M_A)~.
\ee
It should be noticed that this is not the only possible choice, incidentally, it is the %%@
simplest one, 
since our proposal here is not to find the whole parameters range that ensures a net 
attractive potential in both cases ($s$- and $p$-wave), but only to verify that such a 
possibility may indeed be implemented.

\section{Low-energy electron-electron bound states} 
In Section $1$, we showed that, in the model presented here, electrons may attract each 
other in three dimensions through scattering processes where a massive gauge boson 
and a Higgs are involved. This electronic attraction might 
favour a bound state.

The radial Schr\"odinger equation associated to both the electron-electron pairing states, %%@
$s$- and $p$-wave, reads  
\be
\frac{d^2g^{s,p}(r)}{dr^2}+\frac{\frac14-l^2}{r^2}g^{s,p}(r)+2m_b[E_b^{s,p}-\cv^{s,p}(r)]g
^{s,p}(r)=0~,
\ee
where $l$ is the orbital angular momentum and $m_b$ is the reduced effective mass, given by 
\be
m_b=\frac{m}{2} = \frac12 (m_e+\l_{ep}v^2)~.
\ee

By recalling the scattering potentials ${\cv}^{s}$ and ${\cv}^{p}$,
%\ba
%{\cv}^{s}(r)\!\!\!\!&=&\!\!\!\!-\frac{1}{2\p}
%\left[2\l_{ep}^2v^2+\frac{e^2}{\l_c}\right]
%K_0(2v^2\frac{e^2}{\l_c}r)~,\nonumber\\
%{\cv}^{p}(r)\!\!\!\!&=&\!\!\!\!-\frac{1}{2\p}
%\left[2\l_{ep}^2v^2-\frac{e^2}{\l_c}\right]
%K_0(2v^2\frac{e^2}{\l_c}r)~,\nonumber 
%\ea 
it can be concluded that for weak-coupling electrons, where 
$2\l_{ep}^2v^2$$\ll$$\frac{e^2}{\l_c}$, only ${\cv}^{s}$ is attractive, 
therefore, only $s$-type bound states are available.
In the case of strong-coupling electrons, for which 
$2\l_{ep}^2v^2$$\geq$$\frac{e^2}{\l_c}$, in addition to the possible $s$-type bound states 
(since ${\cv}^{p}$ is also attractive), potential $p$-type bound states could appear.  

\subsection{Some very preliminary results}
In the Table $1$, we collect some results we have succesfully applied (they are assigned an %%@
{\bf OK}) to the case of four planar strong-coupling high-$T_c$ superconductors. The first two %%@
cases displayed in the Table $1$ are weak-coupling BCS superconductors. Due to the fact that %%@
BCS  superconductivity effect is not a quasi-planar phenomenon, the model proposed here does %%@
not fit (one assigns an {\bf X}), as it should be expected, the experimental results at all.

%%%%%%%%%%%%%%%%%%%%%%%%%%%%%%%%%%%%%%%%%%%%%%%%%%%%%%%
\begin{table}[tbp]
\par
\begin{center}
\begin{tabular}{|c||c|c|c|c|c|c|c|}
\hline
& $T_c(K)$ & $\l_c({\interior A})$ & $\l_{ep}$ & $2\D(0)(meV)$ & $E_b(meV)$ & $v^2(meV)$  \\ 
\hline\hline
TaS$_2$ & $0,6$ & $48000$ & $0,41$ & $0,182$ & $-0,384$ [{\bf X}] & $\approx 10$  \\ \hline
NbSe$_2$ & $7,1$ & $4800$ & $0,74$ & $2,15$ & $-4,68$ [{\bf X}] & $\approx 0,1$  \\ \hline
YBa$_2$Cu$_3$O$_7$ & $87$ & $1700$ & $2,5$ & $30,0$ & $-30,0$ [{\bf OK}] & $1,05$  \\ \hline
Tl$_2$Ba$_2$Ca$_2$Cu$_3$O$_{10}$ & $105$ & $4800$ & $2,0$ & $28,0$ & $-28,0$ [{\bf OK}]& %%@
$2,10$  \\ \hline
Bi$_2$Sr$_2$CaCu$_2$O$_8$ & $109$ & $5000$ & $2,6$ & $53,4$ & $-53,4$ [{\bf OK}] & $2,73$  \\ %%@
\hline
HgBa$_2$Ca$_2$Cu$_3$O$_8$ & $131$ & $1980$ & $2,5$ & $48,0$ & $-48,0$ [{\bf OK}] & $2,20$  \\ %%@
\hline 
\end{tabular}
\end{center}
\label{table}
\caption{Data we try to fit.}
\end{table} 

\section{Conclusions}
Four high-$T_c$ and two BCS superconductors have been analyzed. In the high-$T_c$ cases, the %%@
model proposed here fits quite well their respective gaps ($2\D(0)$). Contrary, for the 
BCS cases, the model failed, as expected, due to the fact that BCS superconductivity is not a %%@
quasi-planar phenomenon.
It is now under consideration the most general case, $M_H$$\neq$$M_A$, and its possible %%@
application to the re-entrant superconductivity effect \cite{boebinger}. Also, we are %%@
searching for bound states in the Maxwell-Chern-Simons model coupled to QED$_3$ 
\cite{girotti} with spontaneous breaking of the $U(1)$-symmetry. The issue of the 
thermodynamical and statistical properties (phase transitions, specific heats\dots) of a %%@
planar electron gas (for the ideal case, see \cite{blas}) shall be investigated further.

\end{document}